\documentclass[%
twocolumn,
amsmath,amssymb,
prl,
]{revtex4-2}

\usepackage{graphicx}
\usepackage{dcolumn}
\usepackage{bm}
\usepackage{hyperref}

\usepackage{float}
\usepackage{comment}

\newcommand{\bvec}[1]{\bm{#1}}
\newcommand{\unid}[1]{\ensuremath{\:\mathrm{#1}}}

\newcommand{\wRR}{\omega_\text{RR}}

\newcommand{\UE}{U_\mathcal{E}}
\newcommand{\UB}{U_\mathcal{B}}

\newcommand{\uE}{u_\mathcal{E}}
\newcommand{\uB}{u_\mathcal{B}}
\newcommand{\avg}[1]{\left\langle #1 \right\rangle}

\graphicspath{{Imagens/}}

\begin{document}

\preprint{}

\title{A gravitational polarization drift in curved-spacetime guiding-center theory}

\author{João Joaquim}
\email{joao.s.joaquim@tecnico.ulisboa.pt}
\author{Francisco Assunção}%
\author{Luís O. Silva}%
\email{luis.silva@tecnico.ulisboa.pt}
 
\affiliation{GoLP/Instituto de Plasmas e Fusão Nuclear, Instituto Superior Técnico, Universidade de Lisboa, 1049-001 Lisboa, Portugal}

\date{November 30, 2026}

\begin{abstract}

We derive the guiding-center equations of motion with radiation reaction in curved spacetime, in covariant and 3+1 form, for arbitrary field configurations, and validate them against full-orbit integration. 
Our framework enables realistic-parameter simulations of charged-particle motion in strong gravity and compact-object magnetospheres at reduced computational cost.
The gravitationally induced variation in particle inertia generates a new first-order polarization drift that persists in uniform fields, and whose power exceeds that of gravity for relativistic $E\times B$ drifts. 
We show, using the equivalence principle, that this drift belongs to a general class of polarization drifts induced by external forces that change the particle energy, generalizing the $F\times B$ drift. 

\end{abstract}

\maketitle



Charged-particle motion in strong gravity lies at the intersection of general relativity and plasma physics. Understanding the dynamics of charged particles \cite{carter1968,wald1974,spmotionaliev2002,spmotionfrolov2010,dewittbrehme1960} is fundamental to the study of compact-object magnetospheres, where ultra-strong gravitational and electromagnetic fields couple to the collective plasma dynamics \cite{teukolskycompactobjects,limitsextremeplasma,magnetospherereviewphilippov}.
A first-principles description that includes curved spacetime and radiative effects is required to model several key astrophysical phenomena in these environments, such as pair-production gaps \cite{chen2014electrodynamics,crinquand2020BHpair}, coherent emission mechanisms in neutron stars \cite{melrosereview,frbreview,pabloECMIrings}, and black-hole accretion flows and jets \cite{blandfordznajek1977,galishnikovaaccretion}. 
These systems exhibit a very large scale separation, with gyroradii more than ten orders of magnitude smaller than the characteristic system size \cite{bacchini2020}, thereby rendering full-orbit integration and kinetic approaches computationally prohibitive \cite{limitsextremeplasma,paramrescalingPIC}.

The guiding-center (GC) approximation circumvents this scale separation by averaging over the fast particle gyromotion \cite{vandervoort1960,northrop1961,littlejohn1983,carybrizard2009}.
Previous works have extended the GC formalism to curved spacetime either under restrictive assumptions \cite{bacchini2020} or in a covariant form not suited for numerical codes \cite{beklemishev2004,sonstephanovGC2024,trent2025}. Conversely, the inclusion of radiation reaction (RR) in the guiding-center approximation has only been performed in flat spacetime \cite{hirvijoki2015,GCFrancisco2026}. 

In this Letter, we present the first formulation of curved-spacetime guiding-center theory with radiation reaction, and identify a novel gravitational polarization drift which, by the equivalence principle, belongs to a broader class of external-force-induced polarization drifts.

\medskip

The motion of a charged particle of mass $m$, charge $q$, and four-velocity $u^\mu = \frac{d x^\mu}{d\tau}$ in a curved spacetime background, under the action of an external electromagnetic field $F^{\mu\nu}$ and classical radiation reaction, is described by the geodesic equation of motion \cite{mtw1973,landaulifshitzbook}:
\begin{flalign}
    \label{eq:geodesiceq}
      \frac{d u^\mu}{d\tau} = &-\Gamma^\mu_{\alpha \beta} u^\alpha u^\beta + \frac{q}{m} F^{\mu\nu} u_\nu + \frac{2q^3}{3m^2} \Big[ (\nabla_\alpha F^{\mu\nu}) u_\nu u^\alpha \nonumber \\ &+ \frac{q}{m} \left( F^{\mu\nu} F_{\nu\alpha} u^\alpha + F^{\alpha\beta} F_{\beta\nu} u_\alpha u^\nu u^\mu \right) \Big] \ ,
\end{flalign}
\noindent
where $\tau$ is the proper time and $\Gamma^\mu_{\alpha\beta}$ are the Christoffel symbols. The radiation reaction force is modeled according to the Landau-Lifshitz formulation in curved spacetime (see the End Matter for further details).
We employ geometrized CGS units $G = c = 1$, and the mostly-plus metric signature.

The GC approximation decomposes the particle position $x^\mu \approx X^\mu + \rho^\mu$ into a slowly varying guiding-center position $X^\mu$ and a fast gyromotion perpendicular to the magnetic field, $\rho^\mu$, and is formally valid when the system is magnetically dominated ($\mathcal{I}_1 > 0$, defined below) and the electromagnetic field varies slowly in space and time over one particle gyro-orbit \citep{vandervoort1960}. 
Curved spacetime and radiation reaction do not modify the underlying ordering as long as the gravitational force is much weaker than the electromagnetic force \cite{trent2023}, and the energy lost to radiation over one gyroperiod is negligible \cite{GCFrancisco2026}. 
The validity conditions of the GC approximation are thus generally satisfied in highly magnetized systems, such as compact-object magnetospheres, and are stated as \cite{vandervoort1960,trent2023}:
\begin{align}
    \label{eq:GCvalidityEM}
     &\left| \rho^\alpha \partial_\alpha F^{\mu \nu} \right| \ll \left| F^{\mu \nu} \right| \ ; \ \left| \dot{X}^\alpha \partial_\alpha F^{\mu \nu} \right| \ll \left| \omega_c F^{\mu \nu} \right| \ ; \\ 
     \label{eq:GCvalidityRRGR}
     &\left| m\Gamma^\mu_{\alpha \beta} u^\alpha u^\beta \right| \ll \left|q F^{\mu\nu} u_\nu \right| \ ; \
     \wRR \ll |\omega_c| \ ,
\end{align}
\noindent
where the characteristic frequency for radiation damping $\wRR = \frac{2q^4}{3m^3}(\mathcal{B}^2+\mathcal{E}^2)$, and the generalized gyrofrequency $\omega_c = q\mathcal{B}/m$ are determined from the eigenvalues of the electromagnetic tensor \cite{vandervoort1960}:
\begin{align}
     \label{eq:Einvariant}
     \mathcal{E}^2 = \frac{1}{2}\left(\sqrt{\mathcal{I}_1^2 + 4\mathcal{I}_2^2} - \mathcal{I}_1 \right) \ ; \\
     \label{eq:Binvariant} 
     \mathcal{B}^2 = \frac{1}{2}\left(\sqrt{\mathcal{I}_1^2 + 4\mathcal{I}_2^2} + \mathcal{I}_1 \right) \ ,
\end{align}
\noindent
which are in turn defined in terms of the invariants $\mathcal{I}_1 = \frac{1}{2}F^{\mu\nu} F_{\mu\nu} = B^2-E^2$ and $\mathcal{I}_2 =\frac{1}{8}\epsilon^{\mu\nu\alpha\beta} F_{\mu\nu} F_{\alpha\beta} = \bvec{E}\cdot\bvec{B}$ \cite{fradkin1978,li2021RR}. 
We employ the sign convention such that $\mathcal{B} \geq 0$ and $\mathcal{E}\mathcal{B} = \bvec{E}\cdot\bvec{B}$.

Conditions (\ref{eq:GCvalidityEM})-(\ref{eq:GCvalidityRRGR}) are satisfied in the classical regime of radiation reaction, $\chi = \sqrt{(F^{\mu\nu}u_\nu)^2}/B_\text{sc} \ll 1$, with $B_\text{sc} = \frac{m_e^2 c^3}{e \hbar}$ the Schwinger magnetic field \cite{schwingerlimit,quantumRRparam1}, and when the guiding-center expansion parameter is small, $\varepsilon = \left|\rho^\mu \partial_\mu \right| \approx \frac{\gamma m}{|q|\mathcal{B} L} \ll 1$, with $L$ the length scale of variation of the electromagnetic field. 
Expanding the GC four-velocity in the eigenspaces of the electromagnetic tensor yields the evolution equation for the position:
\begin{equation}
    \label{eq:GCXcovariant}
     \frac{dX^\mu}{d\tau} \equiv U^\mu = \UE^{\mu} + \UB^{\mu} \ .
\end{equation}
Since $\mathcal{E}$ and $\mathcal{B}$ are the electric and magnetic field amplitudes in the reference frame where $\bvec{E}\parallel\bvec{B}$, the components $\UE^{\mu}$ and $\UB^{\mu}$ can be interpreted respectively as the generalized guiding-center four-velocities parallel and perpendicular to the magnetic field.

When the guiding-center approximation is valid, the trajectory of a charged particle is accurately described by a gyroaveraged first-order expansion of Eq. (\ref{eq:geodesiceq}) around $x^\mu = X^\mu$ (see the Supplemental Material \cite{suppmaterial} for a detailed derivation). The guiding-center equation of motion in covariant form is then given by: 
\begin{equation}
    \label{eq:GCucovariant}
     \frac{d\UE^\mu}{d\tau} = -\Gamma^\mu_{\alpha \beta} \UE^\alpha \UE^\beta + \frac{q}{m} F^{\mu\nu} U_{\nu} - \frac{\mu}{m} \nabla^\mu \mathcal{B} - \frac{2\wRR\mu\mathcal{B}}{m} \UE^\mu \ ,
\end{equation}
where $\mu = \frac{m\left| \dot{\rho}^\nu \right|^2}{2\mathcal{B}}$ is the magnetic moment of the charged particle \cite{fradkin1978}, and the zeroth-order integrability condition $\left| \UE^\mu \right|^2 = -1 - 2\mu\mathcal{B}/m$ was employed. 
The coefficient of the RR force term is always positive, and thus this interaction acts as a friction force that scales with $\wRR$ and opposes the guiding-center motion, as expected \cite{marija2016RR}.

The perpendicular four-velocity $\UB^\mu$ is uniquely determined from Eq. (\ref{eq:GCucovariant}). Projecting it onto the $\mathcal{B}$-eigenspace and contracting with the electromagnetic tensor gives:
\begin{equation}
    \label{eq:UBcovariant}
     \UB^\mu = -\frac{m\left( \mathcal{E}^2 \delta^{\mu}_{\ \alpha}
     - F^{\mu}_{\ \beta}F^{\beta}_{\ \alpha} \right)}
     {q\mathcal{B}^2\left( \mathcal{E}^2+\mathcal{B}^2 \right)}F^{\alpha}_{\ \nu}
     \left( \UE^\lambda \nabla_\lambda \UE^\nu + \frac{\mu}{m} \nabla^\nu \mathcal{B} \right) .
\end{equation}
Eq. (\ref{eq:UBcovariant}) contains all first-order drift velocities in covariant form \cite{sonstephanovGC2024}, and thus shows that radiation reaction does not introduce any new particle drifts at first order. 
Its effect, however, is manifest in the perpendicular momentum of the particle, which is depleted due to synchrotron emission \cite{landaulifshitzbook}.

The magnetic moment is thus no longer conserved under adiabatic variations of the electromagnetic field \cite{medvedev2026synchrotron,GCFrancisco2026}. Its evolution equation can be derived directly by applying the adiabaticity condition $\frac{d\mathcal{B}}{d\tau} \approx 0$, and reads:
\begin{equation}
    \label{eq:GCmu}
     \frac{d\mu}{d\tau} = -2\wRR \left(1+\frac{2\mu\mathcal{B}}{m} \right)\mu \ .
\end{equation}
Analogously to the last term in Eq. (\ref{eq:GCucovariant}), the source term of Eq. (\ref{eq:GCmu}) is always negative, and therefore the magnetic moment decreases monotonically and non-linearly.
The adiabatic invariance of the magnetic moment is recovered in the limit $\wRR \to 0$, even if curved-spacetime effects are present. This is a consequence of the hierarchy of the guiding-center approximation (\ref{eq:GCvalidityEM})-(\ref{eq:GCvalidityRRGR}).

The system of equations (\ref{eq:GCXcovariant})-(\ref{eq:GCmu}) fully determines the covariant guiding-center variables $(X^\mu, U^\mu, \mu)$, and generalizes Eqs. (A2)-(A3) in Ref. \cite{sonstephanovGC2024} and Eq. (33) in Ref. \cite{trent2025}, to the case in which classical radiation reaction is included. 

\medskip

The physical interpretation of Eqs. (\ref{eq:GCXcovariant})-(\ref{eq:UBcovariant}) is nevertheless obscured by the covariant form.
We resort to the 3+1 formalism of general relativity (GR) \cite{gourgoulhon3+1} to decompose Eqs. (\ref{eq:GCXcovariant})-(\ref{eq:UBcovariant}) into a more physically interpretable form. The derivation is presented in detail in the Supplemental Material \cite{suppmaterial}. 
It can be shown directly from the eigenspace decomposition of the electromagnetic tensor that for any general field configuration with $\mathcal{B}\neq 0$, the parallel guiding-center four-velocity $\UE^\mu$ can be 3+1 decomposed as:
\begin{equation}
    \label{eq:Uk3+1}
     \UE^\mu = \gamma (n^\mu + v_E^\mu) + u_\parallel b_*^\mu \ ,
\end{equation}
where $n^\mu = \left( (\partial_t)^\mu -\beta^\mu \right)/\alpha$ is the four-velocity of fiducial observers, $\alpha = (-g^{00})^{-1/2}$ is the lapse function, and $\beta^i = \alpha^2 g^{0i}$ is the shift vector, which respectively quantify gravitational time dilation and coordinate frame rotation \cite{gourgoulhon3+1}. 

The purely spatial vectors $b_*^\mu$ and $v_E^\mu$ are, respectively, the generalized magnetic field direction, which has unit norm, and the electric drift velocity:
\begin{align}
     \label{eq:bstar}
    \bvec{b}_* = \frac{\mathcal{B} \bvec{B} + \mathcal{E}\bvec{E}}{\gamma_E (\mathcal{B}^2+\mathcal{E}^2)} \ \ ; \ \
     \bvec{v}_E = \frac{\bvec{E}\times\bvec{B}}{\gamma_E^2 (\mathcal{B}^2+\mathcal{E}^2)} \ ,
\end{align}
with $\bvec{v}_E \cdot \bvec{b}_* = 0$. These expressions generalize the special-relativistic ones presented in Refs. \cite{beklemishev1999,bacchini2020}, and reduce to their familiar versions $\bvec{b}_* = \bvec{B}/B$ and $\bvec{v}_E = \bvec{E}\times\bvec{B}/B^2$ when $\bvec{E}\cdot\bvec{B} = 0$, such that $\mathcal{E}=0$ and $\mathcal{B}=B/\gamma_E$. 
In addition, the guiding-center Lorentz factor is:
\begin{equation}
    \label{eq:gammaGC}
     \gamma = \gamma_E\sqrt{1+u_\parallel^2+2\mu\mathcal{B}/m} \ ,
\end{equation}
where $\gamma_E = \sqrt{\frac{B^2+\mathcal{E}^2}{\mathcal{B}^2+\mathcal{E}^2}} = (1-v_E^2)^{-1/2}$ is the Lorentz factor associated with the $\bvec{E}\times\bvec{B}$ drift velocity, and $u_\parallel = U^\mu b _{*\mu}$ is the generalized parallel GC four-velocity component. 

\medskip

The evolution equation for the parallel velocity $u_\parallel$ is derived by substituting (\ref{eq:Uk3+1}) into the spatial projection of (\ref{eq:GCucovariant}) and contracting with $\bvec{b}_*$. This results in: 
\begin{align}
    \label{eq:3+1du//dt}
     \frac{d u_\parallel}{dt} &= - \gamma\frac{D v_E^i}{Dt} \bvec{e}_i \cdot \bvec{b}_* +\frac{\alpha q}{m\gamma_E^2}\bvec{E}\cdot\bvec{b}_* - \frac{\alpha\mu}{\gamma m}\bvec{\nabla}\mathcal{B} \cdot\bvec{b}_* \\ \nonumber & \quad \ -\frac{2\alpha\mu\mathcal{B}}{\gamma m} \wRR u_\parallel + \alpha\gamma \bvec{g}\cdot\bvec{b}_* + \alpha H\cdot\bvec{U}_\mathcal{E}\cdot\bvec{b}_* \ ,
\end{align}
where $\frac{D}{Dt} \equiv \frac{\partial}{\partial t} + \left(\frac{\alpha}{\gamma}\bvec{U}_\mathcal{E} - \bvec{\beta}\right)\cdot\bvec{\nabla}$ is the spatial material covariant derivative along the guiding-center trajectory, and we assume a stationary spacetime metric, $\partial_t g_{\mu\nu} = 0$ \cite{galtsov1978}. 
Our results are also valid for non-stationary spacetimes (see the Supplemental Material for the general case). The spatial coordinate basis is $\{\bvec{e}_i\}$, and the zeroth-order time dilation relation $\frac{d\tau}{dt} = \frac{\alpha}{\gamma}$ was employed. 

Eq. (\ref{eq:3+1du//dt}) provides the general-relativistic generalization of the guiding-center force equation \cite{vandervoort1960,northrop1961}. The source terms on its right-hand side consist of the inertial force associated with changes in the electromagnetic field; the parallel electric-field-powered acceleration; the generalized magnetic mirror force, which decays over time due to radiative dissipation \cite{GCFrancisco2026}; the radiation reaction drag, and the gravitational acceleration and frame-dragging forces. 
The 3+1 decomposed gravitational force is written in terms of the gravitational acceleration, $\bvec{g} = -\bvec{\nabla} \ln{\alpha}$, and the gravitomagnetic tensor $H_{ij} = \alpha^{-1} \nabla_i \beta_j$ and extrinsic curvature tensor $K_{ij} = H_{(ij)}$, which encode the effect of dragging of inertial frames \cite{3+1EM,gourgoulhon3+1}.


The 3+1 system of guiding-center equations is closed by the evolution equation for the guiding-center position. Inserting the spatial projections of Eqs. (\ref{eq:UBcovariant}) and (\ref{eq:Uk3+1}) into Eq. (\ref{eq:GCXcovariant}) results in:
\begin{align}
    \label{eq:dXdtdrifts}
     &\frac{dX^i}{dt} \bvec{e}_i = \alpha \left( \frac{u_\parallel \bvec{b}_*}{\gamma} + \bvec{v}_E \right) - \bvec{\beta} + \frac{m\gamma_E}{q\mathcal{B}} \bvec{b}_* \times \Bigg[ \nonumber u_\parallel \frac{D b_*^i}{Dt}\bvec{e}_i \\ &+ \gamma \frac{D v_E^i}{Dt}\bvec{e}_i + \frac{\alpha\mu}{\gamma m} \bvec{\nabla}\mathcal{B} + \frac{\mu}{\gamma m} \left( \frac{\partial\mathcal{B}}{\partial t} - \bvec{\beta}\cdot\bvec{\nabla}\mathcal{B} \right)\bvec{v}_E \\ \nonumber &-\alpha\gamma\bvec{g} - \alpha H\cdot\bvec{U}_\mathcal{E} + \left( \alpha\bvec{g}\cdot\bvec{U}_\mathcal{E} + \frac{\alpha}{\gamma} K\cdot\bvec{U}_\mathcal{E}\cdot\bvec{U}_\mathcal{E} \right)\bvec{v}_E \Bigg] \ .
\end{align}
Together with the time-dilated Eq. (\ref{eq:GCmu}) and Eq. (\ref{eq:3+1du//dt}), Eq. (\ref{eq:dXdtdrifts}) forms a closed dynamical system that constitutes the general-relativistic guiding-center equations of motion including radiation reaction in 3+1 form, valid in any coordinate system.
We emphasize that our formulation is not restricted to nearly crossed electric and magnetic fields, and thus does not involve any apparent drifts resulting from a small parallel electric field expansion $\mathcal{E} \ll \mathcal{B}$, generalizing previous works \cite{vandervoort1960,northrop1961,bacchini2020}. Such terms are contained in the generalized parallel vector $\bvec{b}_*$. 

The guiding-center velocity decomposes into a leading-order term composed of the generalized parallel velocity, the gravitationally redshifted electric drift and the shift vector, plus all first-order drift velocities, as seen in Eq. (\ref{eq:dXdtdrifts}).
These are, from left to right, the general-relativistic extensions of the curvature, polarization and gradient drifts \cite{vandervoort1960}, the relativistic $\partial \mathcal{B}/\partial t$ drift \cite{northrop1961,ripperda2017GC1}, to which we will refer from now on as the mirror polarization drift, and the external-force drifts for the gravitational acceleration and the frame-dragging forces.
The radiation reaction force does not give rise to any particle drifts at first order \cite{GCFrancisco2026}, as expected from Eq. (\ref{eq:UBcovariant}), and causes the gradient and mirror polarization drifts to vanish due to the magnetic moment depletion. All other first-order drift velocities, which do not depend directly on $\mu$, are only reduced through the radiative decrease of $u_\parallel$ and $\gamma$, and survive once the perpendicular momentum has been radiated.

\medskip

\begin{figure*}[t]
    \centering
    \includegraphics[width=\linewidth]{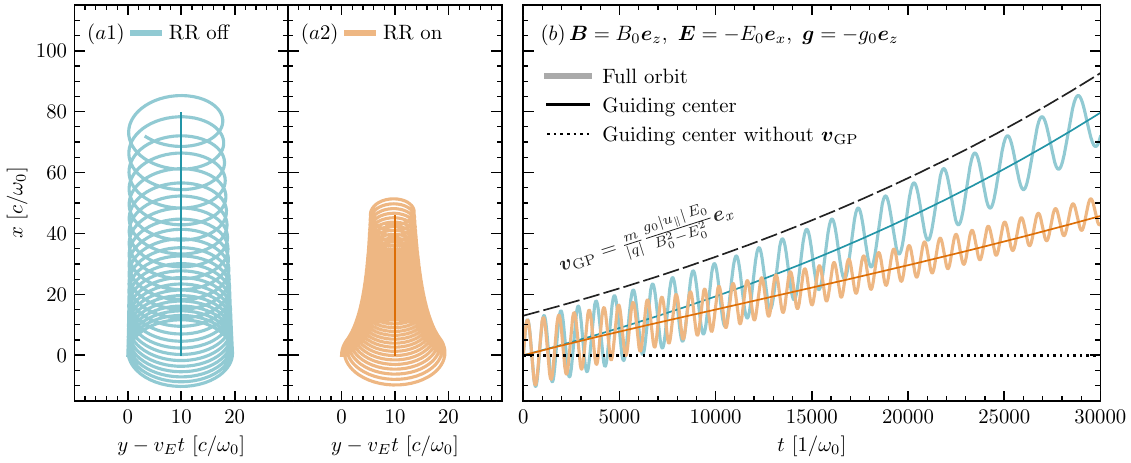}
    \caption{Gravitational polarization drift trajectories of an electron initialized with $u_\perp = 10$ and $u_\parallel = -100$ in the electric-drift frame, in uniform perpendicular electric and magnetic fields with $E_0/B_0 = 0.5$, subject to a uniform gravitational acceleration $g_0 = 2.5\times10^{-5}\,c\,\omega_0$ antiparallel to $\bvec{B}$, with and without radiation reaction, for $\omega_\text{RR}/\omega_0 = 7.5\times10^{-5}$. The full-orbit integration (see the End Matter) must resolve the gyromotion, whereas the guiding-center integration does not. Simulation parameters are normalized to $\omega_0 = |q|B_0/mc$.} 
    \label{fig:GPdrift}
\end{figure*}

Surprisingly, the 3+1 formulation of the curved-spacetime guiding-center equations reveals a general-relativistic drift velocity which was not identified in previous works \cite{bacchini2020,sonstephanovGC2024,trent2025}, the last term in Eq. (\ref{eq:dXdtdrifts}). This drift is charge-dependent, enters the guiding-center expansion at first order and persists in spatially uniform fields.
%
Its origin is the gravitationally induced variation in particle energy, $\left(\frac{d\gamma}{dt}\right)_\text{GR} = \alpha\bvec{g}\cdot\bvec{U}_\mathcal{E} + \frac{\alpha}{\gamma} K\cdot\bvec{U}_\mathcal{E}\cdot\bvec{U}_\mathcal{E}$, which produces an effective inertial force along the electric drift, of the form $\bvec{F}_\text{eff} = -m\left(d\gamma/dt\right)_\text{GR} \bvec{v}_E$. 
This force subsequently generates a perpendicular drift velocity distinct from the usual gravitational-force drifts, and directed along $\bvec{E}_\perp = \gamma_E\mathcal{B}(\bvec{b}_*\times\bvec{v}_E)$, which is given by:
\begin{equation}
    \label{eq:vGP}
     \bvec{v}_\text{GP} = \frac{m}{q\mathcal{B}^2} \left(\frac{d\gamma}{dt}\right)_\text{GR} \bvec{E}_\perp \ .
\end{equation}
This new perpendicular motion is the consequence of inertial corrections to the electric-drift frame resulting from gravitational effects, and thus we identify it as a gravitational, or general-relativistic, polarization (GP) drift.
In Fig. \ref{fig:GPdrift}, we show the effect of the GP drift on single-particle motion in uniform fields $\bvec{g}\parallel-\bvec{B}\perp\bvec{E}$, where the only additional drift that persists is the $\bvec{E}\times\bvec{B}$ drift. The full-orbit motion is reproduced by the guiding-center equations only when $\bvec{v}_\text{GP}$, given by Eq. (\ref{eq:vGP}), is included.
%

To estimate the importance of the general-relativistic drift, we consider a non-rotating spacetime configuration. Given that $\left|\bvec{E}_\perp \right| = \mathcal{B}\gamma_E v_E$ and $|\bvec{U}_\mathcal{E}| \sim \gamma$, the characteristic magnitudes of the $\bvec{g}\times \bvec{B}$ and gravitational polarization drifts are respectively $v_\text{G} = \frac{\alpha\gamma_E\gamma g}{\left|\omega_c\right|}$ and $v_\text{GP} \sim \frac{\alpha\gamma_E\gamma g v_E}{\left|\omega_c\right|}$. We obtain the following scaling formulae for the drift magnitudes:
\begin{align}
     \label{eq:vGPmag}
     \frac{v_\text{GP}}{v_\text{G}} \sim \frac{v_E}{c} \ ; \ \frac{v_\text{GP}}{v_E} \sim \gamma_E\gamma\frac{\omega_\text{GR}}{\left|\omega_c\right|} = \frac{\gamma_E\gamma m g }{|q|\mathcal{B}} \ ,
\end{align}
where the speed of light was restored for physical clarity and $\omega_\text{GR} = g/c$ is the characteristic frequency scale for gravitational effects. We note that in rotating spacetimes, the frame-dragging frequency $\omega_\text{FD} \sim c \sqrt{K_{ij}K^{ij}}$ satisfies $\omega_\text{FD} \lesssim \omega_\text{GR}$ \cite{bardeen1972} and thus the parameter $\omega_\text{GR}$ can still be used as an upper limit in settings where $K_{ij} \neq 0$, making the scale dependences of Eq. (\ref{eq:vGPmag}) general. 
These scalings show that the typical drift ordering is $v_E \gg v_\text{G} \gtrsim v_\text{GP}$, in accordance with the GC expansion hierarchy, and confirm the relevance of the gravitational polarization drift in relativistic astrophysical settings where gravity is strong enough that the standard $\bvec{g}\times\bvec{B}$ drift must be accounted for. This is commonly the case in compact-object magnetospheres \cite{bacchini2020,trent2023}.

\medskip

All first-order drift velocities, including the gravitational polarization drift, are kinematically negligible with respect to the electric drift in highly magnetized environments \cite{bacchini2020}.
Nevertheless, as the electric drift performs no work, $\bvec{v}_E\cdot\bvec{E}=0$, and our GC expansion allows for a zeroth-order perpendicular electric field, the electric work done upon the first-order drifts enters the energy balance at the same order as gravity, and thus becomes of leading order in force-free magnetospheres, where $\bvec{E}\cdot\bvec{B}=0$.
The guiding-center energy equation is the temporal projection of Eq. (\ref{eq:GCucovariant}):
\begin{flalign}
     \label{eq:3+1dgammadt}
     \frac{d \gamma}{dt} &= \frac{\alpha q u_\parallel}{m\gamma} \bvec{E}\cdot\bvec{b}_{*} + \frac{\alpha q}{m\gamma}\bvec{U}_\mathcal{B}\cdot \bvec{E} + \frac{\mu}{\gamma m} \left( \frac{\partial\mathcal{B}}{\partial t} - \bvec{\beta}\cdot\bvec{\nabla} \mathcal{B} \right) \nonumber \\ & -\frac{2\alpha\mu\wRR\mathcal{B}}{m} + \alpha \bvec{g}\cdot\bvec{U}_\mathcal{E} + \frac{\alpha}{\gamma} K\cdot\bvec{U}_\mathcal{E}\cdot\bvec{U}_\mathcal{E} \ ,
\end{flalign}
and accounts for the work done, from left to right respectively, by the electric field along the parallel direction and first-order drift velocities, mirror force, radiation reaction, gravitational acceleration and frame-dragging.
The flat-spacetime reduction of Eq. (\ref{eq:3+1dgammadt}), with $\alpha=1$, $\bvec{\beta}=\bvec{g}=K=0$ \cite{vandervoort1960,northrop1961}, can be employed to identify particle energization mechanisms in kinetic simulations \cite{dahlin2014energization,guo2014energization}, an analysis that has been performed only qualitatively in curved spacetime \cite{parfreyjets,crinquand2020BHpair,galishnikovaaccretion,vos2025energization}. Eq. (\ref{eq:3+1dgammadt}) therefore allows for a detailed study of the contribution of each separate energy channel.

The power delivered by the gravitational polarization drift is contained in the $\bvec{U}_\mathcal{B}\cdot \bvec{E}$ term:
\begin{align}
     \label{eq:dgammadtGP}
     \left( \frac{d\gamma}{dt} \right)_\text{GP} &\equiv \frac{q}{m}\bvec{v}_\text{GP}\cdot\bvec{E} = \gamma_E^2 v_E^2 \left(\frac{d\gamma}{dt}\right)_\text{GR} \ ,
\end{align}
and therefore supplies an additional fraction $\gamma_E^2 v_E^2$ of the energy exchanged through the gravitational force, exceeding it for $v_E > c/\sqrt{2}$, i.e., for a relativistic electric drift in strong gravity, a setting commonly found in black-hole magnetospheres \cite{blandfordznajek1977,gralla2014forcefree}.
The existence of such a drift is therefore required for energy balance consistency: in the uniform-field setup of Fig. \ref{fig:GPdrift} without RR, both $\gamma_E$ and $\mu$ are constants of motion, such that Eq. (\ref{eq:gammaGC}) differentiates to $d\gamma/dt = \alpha \gamma_E^2 u_\parallel \bvec{g}\cdot\bvec{b}_*$ and matches Eq. (\ref{eq:3+1dgammadt}) only if the GP drift is accounted for.
Without it, the electric field performs no work, Eq. (\ref{eq:3+1dgammadt}) underestimates the energy gain by a factor $\gamma_E^2$ (25\% in Fig. \ref{fig:GPdrift}), and the guiding-center position does not vary along $\bvec{E}$ (the dotted line in Fig. \ref{fig:GPdrift}), violating Noether's theorem: the dynamics is invariant under translations along $\bvec{v}_E$, and therefore the conservation of the gyroaveraged canonical momentum, $P_y = \gamma m (v_E)_y + qBX_x = \text{const}$, requires the guiding center to be displaced along the electric field by $\Delta X_x = \frac{m\Delta \gamma}{qB^2}E_\perp$ whenever its energy changes.
Such displacement leads to the formation of a plasma polarization field $\bvec{P} = \frac{\rho_0\langle\gamma\rangle}{B^2}\bvec{E}_\perp$, where $\rho_0 = \sum_s n_s m_s$ is the mass density and $\langle\cdot\rangle$ denotes a mass-weighted average.

As gravity changes the inertia of the plasma, the corresponding change in polarization is carried by the gravitational polarization current:
\begin{align}
     \label{eq:JGP}
     \bvec{J}_\text{GP} = \frac{\rho_0}{\mathcal{B}^2}\left\langle\frac{d\gamma}{dt}\right\rangle_\text{GR}\bvec{E}_\perp \ .
\end{align}
This is the general-relativistic analog of the classical polarization current $\bvec{J}_\text{P} = \frac{\rho_0\langle\gamma\rangle}{B^2}\frac{\partial\bvec{E}_\perp}{\partial t}$, which accounts for the change in plasma inertia due to a time-varying electric drift.
In a uniform plasma subject to a uniform magnetic field, the polarization field thus evolves as $\partial_t\bvec{P} = \bvec{J}_\text{P} + \bvec{J}_\text{GP}$.
In the absence of the GP drift, charge conservation would require the free charge density $\rho_f$ to evolve at the rate $\partial_t\rho_f + \bvec{\nabla}\cdot\bvec{J}_f = \bvec{\nabla}\cdot\bvec{J}_\text{GP}$, where $\bvec{J}_f$ is the free current density. This implies that gravity would unphysically create free charge simply by changing the plasma inertia.
Moreover, the gravitational polarization current (\ref{eq:JGP}) is in phase with the electric field and can thus exchange energy with electrostatic perturbations. We therefore expect it to modify the growth of Rayleigh-Taylor interchange modes in gravitationally stratified systems \cite{kruskalschwarzschild}.


We further observe that the Einstein equivalence principle states that gravity is locally indistinguishable from reference frame acceleration, and therefore suggests that the gravitational polarization drift presented here belongs to a broader class of drift velocities. 
Any external force per unit mass $F^\mu$ entering Eq. (\ref{eq:GCucovariant}) at first order will generate a drift velocity consisting of the standard $\bvec{F}\times\bvec{B}$ drift, and a corresponding relativistic polarization drift:
%
%
\begin{equation}
    \label{eq:vFext}
     \bvec{v}_d = \frac{\alpha \gamma_E m}{\gamma q\mathcal{B}} \bvec{F}\times\bvec{b}_* + \frac{m}{q\mathcal{B}^2}\left(\frac{d\gamma}{dt}\right)_{\!F}\bvec{E}_\perp \ ,
\end{equation}
where the power delivered by the external force in the fiducial frame is $\left( \frac{d\gamma}{dt}\right)_F = -\frac{\alpha}{\gamma} n_\mu F^\mu$.
Besides gravity, the simplest example is the covariant mirror force $F^\mu=-\frac{\mu}{m} \nabla^\mu \mathcal{B}$, which gives rise to the gradient drift, $\bvec{v}_{\nabla\mathcal{B}} = \frac{\alpha\gamma_E\mu}{\gamma q\mathcal{B}}\,\bvec{b}_*\times\bvec{\nabla}\mathcal{B}$, and to the mirror polarization drift, $\bvec{v}_\text{MP} = \frac{m}{q\mathcal{B}^2}\left(\frac{d\gamma}{dt}\right)_\text{MP}\bvec{E}_\perp$, in Eq. (\ref{eq:dXdtdrifts}).

This class of relativistic polarization drifts (the second term in Eq. (\ref{eq:vFext})) is suppressed in the non-relativistic limit, $v_E \ll c$ and $\left( d\gamma/dt\right)_F \to 0$, and vanishes only for external forces that produce no power in the fiducial frame. This is the case for a gravitational force directed along the perpendicular electric field in a static spacetime, $\bvec{g}\parallel\bvec{E}_\perp$, such that $\left( d\gamma/dt\right)_{\rm GR} = \alpha\bvec{g}\cdot\bvec{U}_\mathcal{E}=0$, for example.

Conversely, the external forces that deliver power without generating any drifts are those whose spatial part in the electric-drift frame is directed along $\bvec{b}_*$, i.e., $\frac{\alpha}{\gamma} \bvec{F}-\left( \frac{d\gamma}{dt}\right)_F \bvec{v}_E \parallel \bvec{b}_*$. Physically, the perpendicular momentum supplied by these forces is exactly that required for the particle to remain in the electric-drift frame as its energy changes, so that the terms in Eq. (\ref{eq:vFext}) cancel.
One such example is the radiation reaction force, whose gyroaverage lies in the $\mathcal{E}$-eigenspace and is thus annihilated by the $\mathcal{B}$-eigenspace projector in Eq. (\ref{eq:UBcovariant}), generating no drift velocities at first order.

\medskip

The guiding-center formalism developed here enables the study of charged-particle motion in strong gravity at realistic parameters, and can be directly implemented into state-of-the-art GRMHD \cite{grmhdcomparison2019,trent2026hybrid} or GRPIC \cite{osirisgr,entitygrpic} codes to perform simulations of compact-object magnetospheres at reduced computational cost.
As an example, in the Supplemental Material \cite{suppmaterial} we apply it to realistic-parameter particle trajectories in the jet funnel of M87${}^*$, modeled as a Blandford-Znajek magnetosphere of a Kerr black hole \cite{blandfordznajek1977}.

Furthermore, our guiding-center equations (\ref{eq:GCXcovariant})-(\ref{eq:GCmu}) are valid for arbitrary field geometries and self-consistently include radiation reaction. They thus provide the single-particle dynamics required for the synthetic calculation of synchrotron and curvature radiation spectra, and for extending the gyrokinetic and drift-kinetic framework developed in Refs. \cite{schekochihinlecturenotes,sonstephanovGC2024,kineticGR2026} to relativistic radiative plasmas in curved spacetime. This will be presented in a future publication.

\medskip

\begin{acknowledgments}

We would like to thank P. Bilbao, T. Grismayer, P. Kempski, M. Medvedev, and A. Wierzchucka for discussions. This work was partially supported by the Foundation for Science and Technology, I.P. (FCT) through the project X-Maser (No. 2022.02230.PTDC). 
Simulations were performed at MareNostrum 5 (Barcelona Supercomputing Center, Spain), funded by the FCT project No. 2025.09524.CPCA.A3.
IPFN's activities were supported by FCT through the project UID/50010/2025 (https://doi.org/10.54499/UID/50010/2025).
All data needed to evaluate the conclusions are present in the paper.
\end{acknowledgments}


\makeatletter
\def\bibsection{%
 \par
 \onecolumngrid@push
 \begingroup
  \baselineskip26\p@
  \bib@device{\textwidth}{245.5\p@}%
 \endgroup
 \nobreak\@nobreaktrue
 \addvspace{19\p@}%
 \par
 \onecolumngrid@pop
}%
\makeatother

\bibliography{paper}

\onecolumngrid
\vspace{\baselineskip}

\begin{center}\textbf{\large End Matter}\end{center}
\vspace{0.5\baselineskip}
\twocolumngrid

\appendix

\section{Full-orbit equation of motion with radiation reaction}
\label{app:3+1geodesiceq}

\setcounter{equation}{0}\renewcommand{\theequation}{A\arabic{equation}}

The full-orbit trajectories in Fig. \ref{fig:GPdrift} are obtained by integrating the 3+1-decomposed equations of motion (\ref{eq:geodesiceq}), which are given by \cite{3+1EM,gourgoulhon3+1}:
\begin{flalign}
     \label{eqEM:3+1eqmovx}
     &\frac{dx^i}{dt} \bvec{e}_i = \frac{\alpha\bvec{u}}{\gamma} - \bvec{\beta} \ ; && \\
     \label{eqEM:3+1eqmovu}
     &\frac{D u^i}{Dt}\bvec{e}_i = \alpha\left[ \gamma\bvec{g} + H \cdot \bvec{u} + \frac{q}{m} \left(\bvec{E} + \frac{\bvec{u}}{\gamma} \times \bvec{B} \right) \right] + \frac{\bvec{F}_\text{RR}}{m} \ , &&
\end{flalign}
where the spatial material covariant derivative $D/Dt$ now acts along the full-orbit trajectory, $\gamma = \sqrt{1+\left|\bvec{u}\right|^2}$ is the full-orbit Lorentz factor, and $\bvec{F}_\text{RR}$ is the spatial part of the general-relativistic Landau-Lifshitz radiation reaction force \cite{dewittbrehme1960,landaulifshitzbook,joaquim2026} (see Ref. \cite{RRcurvedspace} for a review of radiation reaction in curved spacetime), included in Eq. (\ref{eq:geodesiceq}):
\begin{align}
    \label{eqEM:curvedspaceLLRR}
    F_\text{RR}^\mu  = \frac{2q^3}{3m} \bigg[ &(\nabla_\alpha F^{\mu}_{\ \nu} ) u^\nu u^\alpha + \frac{q}{m} ( F^{\mu\nu} F_{\nu\alpha} u^\alpha \\ &+ F^{\alpha \beta} F_{\beta\nu} u_\alpha u^\nu u^\mu) \bigg] \ \nonumber . 
\end{align}
The covariant derivative of the electromagnetic tensor is of second order and therefore does not enter the guiding-center expansion at first order.
For completeness, we include it nevertheless in the full-orbit equations of motion (\ref{eqEM:3+1eqmovu}) employed for the numerical validation of the guiding-center equations \cite{suppmaterial}. 

The spatial projection of the Landau-Lifshitz force in curved spacetime (\ref{eqEM:curvedspaceLLRR}) can be shown to be given by:
\pagebreak
\begin{align}
    \label{eqEM:LL3D}
     \bvec{F}_\text{RR} = &\frac{2q^2}{3}\Bigg[ \frac{D F_L^i}{Dt}\bvec{e}_i - \frac{\alpha q}{\gamma m}(\bvec{E}\cdot\bvec{u})\left(\gamma\bvec{g} + K\cdot\bvec{u}\right) \\ \nonumber  &- \alpha\left(H-K\right)\cdot\bvec{F}_L - \frac{\alpha \bvec{u}}{\gamma} \left(F_L^2 - \frac{q^2}{m^2}\left(\bvec{u}\cdot\bvec{E}\right)^2\right) \Bigg]\ ,
\end{align}
where $\bvec{F}_L = \frac{q}{m}\left(\gamma\bvec{E} + \bvec{u}\times\bvec{B}\right)$ is the Lorentz force per unit mass in proper time. Eqs. (\ref{eqEM:3+1eqmovu}) and (\ref{eqEM:LL3D}) are valid in any spacetime, including non-stationary metrics. 

\medskip
Eq. (\ref{eqEM:LL3D}) can be separated as $\bvec{F}_\text{RR} = \bvec{F}_\text{RR}^\text{(LLR)} + \bvec{F}_\text{RR}^\text{(G)}$, where the first term is the reduced Landau-Lifshitz force \citep{landaulifshitzbook}: 
\begin{align}
    \label{eqEM:LLR}
     \bvec{F}_\text{RR}^\text{(LLR)} = &\frac{2\alpha q^4}{3\gamma m^2} \Bigg[ \left(\gamma\bvec{E} + \bvec{u} \times \bvec{B}\right) \times \bvec{B} + \bvec{E}\left(\bvec{u} \cdot \bvec{E}\right) \nonumber && \\ & -\bvec{u} \left(\left(\gamma\bvec{E} + \bvec{u} \times \bvec{B}\right)^2 - \left(\bvec{u} \cdot \bvec{E}\right)^2 \right) \Bigg] \ , &&
\end{align}
and the second is the curved spacetime contribution, arising from the covariant derivative of the electromagnetic tensor in Eq. (\ref{eqEM:curvedspaceLLRR}), which reads, for a stationary metric:
\begin{align}
    \label{eqEM:RRG}
     &\bvec{F}_\text{RR}^\text{(G)} = \frac{2q^3}{3m}\Bigg[ \gamma \frac{DE^i}{Dt} \bvec{e}_i + \bvec{u} \times \frac{DB^i}{Dt}\bvec{e}_i + \alpha \bvec{E}(\bvec{g}\cdot\bvec{u}) \\ \nonumber & + \frac{\alpha}{\gamma}\bvec{E} (K\cdot\bvec{u}\cdot\bvec{u}) + \alpha\gamma \bvec{g}\times\bvec{B} + \alpha (H\cdot\bvec{u})\times\bvec{B} - \alpha\bvec{g}(\bvec{E}\cdot\bvec{u}) \\ & - \frac{\alpha}{\gamma}(K\cdot\bvec{u})(\bvec{E}\cdot\bvec{u}) -\alpha\left(H - K \right)\cdot\left(\gamma\bvec{E}+\bvec{u}\times\bvec{B} \right)  \Bigg] \nonumber \ .
\end{align}
This expression is valid in any coordinate system, and thus fully generalizes Eq. (6) in Ref. \cite{joaquim2026}.

\clearpage
\onecolumngrid

\setcounter{equation}{0}\renewcommand{\theequation}{S\arabic{equation}}\renewcommand{\theHequation}{S\arabic{equation}}
\setcounter{figure}{0}\renewcommand{\thefigure}{S\arabic{figure}}\renewcommand{\theHfigure}{S\arabic{figure}}
\setcounter{section}{0}\renewcommand{\thesection}{S\Roman{section}}

\begin{center}
{\large\bfseries Supplemental Material\par}
\vspace{0.5\baselineskip}
{\bfseries A gravitational polarization drift in curved-spacetime guiding-center theory\par}
\vspace{0.75\baselineskip}
{\small João Joaquim, Francisco Assunção, and Luís O. Silva\par}
\end{center}
\vspace{\baselineskip}

\section*{Derivation of the guiding-center equations of motion}

We employ geometrized units $G = c = 1$, the mostly-plus metric signature, and the Einstein summation convention. Greek and Latin indices denote four- and three-dimensional quantities respectively, spatial quantities are typeset in bold, and their magnitudes and index components in regular typeface. All physical quantities are presented in the CGS system of units.

\subsection*{Covariant form}

The guiding-center equation of motion (\ref{eq:GCucovariant}) is derived by averaging the first-order expansion of Eq. (\ref{eq:geodesiceq}) around $x^\mu = X^\mu$ over one gyration period. 
The particle position is decomposed as $x^\mu = X^\mu + \rho^\mu$, such that the electromagnetic field tensor is Taylor expanded to first order in $\varepsilon = \left|\rho^\mu \partial_\mu \right|$ around $x^\mu \approx X^\mu$ as:
\begin{equation}
    \label{eqSM:Fexpansion}
     F^\mu_{\ \nu}(x) \simeq F^\mu_{\ \nu}(X) + \rho^\alpha\partial_\alpha F^\mu_{\ \nu}(X) + \mathcal{O}\left( (\rho^\alpha \partial_\alpha)^2 \right) \ .
\end{equation}
The four-velocity expansion is simpler in the eigenspaces of the electromagnetic tensor, 
\begin{equation}
    \label{eqSM:uexpansion}
     u^\mu = \uE^{(0)\mu} + \uB^{(0)\mu} + \uB^{(1)\mu} + \mathcal{O}\left( (\rho^\alpha \partial_\alpha)^2 \right) \ ,
\end{equation}
from which we define the gyromotion four-velocity $\dot{\rho}^\mu \equiv u_\mathcal{B}^{(0)\mu}$ and the guiding-center four-velocity components $\UE^\mu \equiv \uE^{(0)\mu}$ and $\UB^\mu \equiv \uB^{(1)\mu}$.
The subspace-decomposed four-velocity components thus satisfy $F^{\mu\alpha} F_{\alpha \nu} u_\mathcal{E}^\nu = \mathcal{E}^2 u_\mathcal{E}^\mu$ and $F^{\mu\alpha} F_{\alpha \nu} u_\mathcal{B}^\nu = -\mathcal{B}^2 u_\mathcal{B}^\mu$.
Inserting the expansions (\ref{eqSM:Fexpansion}) and (\ref{eqSM:uexpansion}) in Eq. (\ref{eq:geodesiceq}), and noting that gyroaveraging makes all terms proportional only to $\rho^\mu$ or its time derivative vanish, leads to:
\begin{align}
    \label{eqSM:GCuwithgyroaverage}
     \frac{d\UE^\mu}{d\tau} = -\Gamma^\mu_{\alpha \beta} \UE^\alpha \UE^\beta + \frac{q}{m} F^{\mu}_{\ \nu} \left( \UE^\nu + \UB^\nu \right) +\omega_\text{RR} \left(1+\left|\UE^\mu \right|^2 \right)\UE^\mu + \left< \frac{q}{m}\rho^\alpha \partial_\alpha F^{\mu}_{\ \nu} \dot{\rho}^{\nu} - \Gamma^\mu_{\alpha \beta} \dot{\rho}^{\alpha} \dot{\rho}^{\beta} \right> \ .
\end{align}
The covariant derivative term in the radiation reaction force is of second order and is therefore neglected.

All terms in Eq. (\ref{eqSM:GCuwithgyroaverage}) are already in final form except for the gyroaverage. This requires the determination of the zeroth-order expressions for $\rho^\mu$ and $\dot{\rho}^\mu$, which we take to be the exact solution for charged-particle motion in a locally uniform, static electromagnetic field including radiation reaction.
Again dropping the covariant derivative term of Eq.~(\ref{eq:geodesiceq}), and the Christoffel symbols, the zeroth-order equation of motion decouples into a separate equation for each eigenspace, such that:
\begin{equation}
    \label{eqSM:drhodt}
     \frac{d \dot{\rho}^\mu}{d\tau} = \frac{q}{m}F^\mu_{\ \nu}\dot{\rho}^\nu - \omega_\text{RR}\left( 1+\left|\dot{\rho}^\mu\right|^2 \right)\dot{\rho}^\mu \ ; \quad 
     \frac{d \left|\dot{\rho}^\mu\right|}{d\tau} = -\omega_\text{RR}\left( 1+\left|\dot{\rho}^\mu\right|^2 \right)\left|\dot{\rho}^\mu\right| \ ,
\end{equation}
whose solutions are, with a nought denoting the initial value,
\begin{equation}
    \label{eqSM:exactsolutionrho}
     \dot{\rho}^\mu = \frac{\dot{\rho}_0^\mu \cos\left(\omega_c\tau\right) + v_{\mathcal{B}0}^\mu \sin\left(\omega_c\tau\right)}{ \sqrt{-\left( \left|\dot{\rho}_0^\mu\right|^2 + \left|U_{\mathcal{E}0}^\mu \right|^2 e^{2\omega_\text{RR}\tau} \right)}} \ ; \quad
     \rho^\mu \simeq \frac{\dot{\rho}_0^\mu\sin\left(\omega_c\tau\right) - v_{\mathcal{B}0}^\mu \cos\left(\omega_c\tau\right) }{\omega_c \sqrt{-\left( \left|\dot{\rho}_0^\mu\right|^2 + \left|U_{\mathcal{E}0}^\mu \right|^2 e^{2\omega_\text{RR}\tau} \right)}}  \ ,
\end{equation}
\noindent
where $v_{\mathcal{B}}^\mu\equiv\frac{1}{\mathcal{B}}F^\mu_{\ \nu}\dot{\rho}^\nu$ and the amplitude of $\rho^\nu$ was assumed to be constant over a gyroperiod, since $\omega_\text{RR} \ll |\omega_c|$.
Substituting (\ref{eqSM:exactsolutionrho}) into the gyroaverage of Eq. (\ref{eqSM:GCuwithgyroaverage}) yields: 
\begin{flalign}
    \label{eqSM:gyroaverage}
     \avg{ \frac{q}{m} \rho^\alpha\partial_\alpha F^\mu_{\ \nu}\dot{\rho}^\nu - \Gamma^\mu_{\alpha\beta}\dot{\rho}^\alpha \dot{\rho}^\beta } &= 
     \frac{\left|\dot{\rho}^\mu\right|^2}{\left|\dot{\rho}^\mu_0\right|^2}\left[ \frac{1}{2\mathcal{B}}\left( \dot{\rho}_0^\alpha\partial_\alpha F^\mu_{\ \nu}v_{\mathcal{B}0}^\nu - v_{\mathcal{B}0}^\alpha\partial_\alpha F^\mu_{\ \nu} \dot{\rho}_0^\nu \right) 
     - \frac{1}{2}\Gamma^\mu_{\alpha\beta}\left( \dot{\rho}_0^\alpha \dot{\rho}_0^\beta + v_{\mathcal{B}0}^\alpha v_{\mathcal{B}0}^\beta \right) \right] && \\
     \label{eqSM:gyroaverageline2}
     &= \frac{\left|\dot{\rho}^\mu\right|^2}{\left|\dot{\rho}^\mu_0\right|^2} \frac{1}{2\mathcal{B}}\left\{ \dot{\rho}_0^\alpha\nabla_\alpha F^\mu_{\ \nu}v_{\mathcal{B}0}^\nu - v_{\mathcal{B}0}^\alpha\nabla_\alpha F^\mu_{\ \nu}\dot{\rho}_0^\nu \right\}  
     = -\frac{\left|\dot{\rho}^\mu\right|^2}{2\mathcal{B}}\nabla^\mu\mathcal{B} \equiv -\frac{\mu}{m} \nabla^\mu\mathcal{B} \ , &&
\end{flalign}
where the homogeneous Maxwell equation $\nabla_\alpha F^\mu_{\ \nu} = -\nabla^\mu F_{\nu\alpha} - \nabla_\nu F_{\alpha}^{\ \mu}$ was employed in step (\ref{eqSM:gyroaverageline2}), and the magnetic moment is defined as $\mu \equiv \frac{m\left|\dot{\rho}^\nu\right|^2}{2\mathcal{B}}$. 
The covariant guiding-center equation of motion is obtained from the zeroth-order integrability condition $\left|U_\mathcal{E}^\mu\right|^2 + \left|\dot{\rho}^\mu\right|^2=-1$, which implies $\left|U_\mathcal{E}^\mu\right|^2 = -1-2\mu\mathcal{B}/m$, leading to:
\begin{equation}
    \label{eqSM:GCucovariant}
     \frac{dU_\mathcal{E}^\mu}{d\tau} = -\Gamma^\mu_{\alpha\beta}U_\mathcal{E}^\alpha U_\mathcal{E}^\beta + \frac{q}{m}F^{\mu\nu}U_\nu - \frac{\mu}{m}\nabla^\mu\mathcal{B} - \frac{2\omega_\text{RR}\mu\mathcal{B}}{m}U_\mathcal{E}^\mu \ .
\end{equation}

The system is closed by Eqs. (\ref{eq:GCXcovariant}) and (\ref{eq:UBcovariant}), and by the evolution equation for the magnetic moment, which is determined by differentiating its definition directly with respect to proper time:
\begin{equation}
    \label{eqSM:muevolution}
     \frac{d\mu}{d\tau} = \frac{m}{2}\left( \frac{2\left|\dot{\rho}^\mu\right|}{\mathcal{B}} \frac{d \left|\dot{\rho}^\mu\right|}{d\tau} - \frac{\left|\dot{\rho}^\mu\right|^2}{\mathcal{B}^2} \frac{d\mathcal{B}}{d\tau} \right) = -2\omega_\text{RR} \left( 1 + \frac{2\mu\mathcal{B}}{m} \right) \mu \ ,
\end{equation}
where the adiabaticity condition $\frac{d\mathcal{B}}{d\tau}\approx 0$ and Eq. (\ref{eqSM:drhodt}) were used.

\subsection*{3+1 decomposition of the parallel guiding-center four-velocity}

The 3+1 formalism foliates four-dimensional spacetime into a set of spacelike hypersurfaces that vary continuously along a global time coordinate $t$ and possess a single unit timelike normal vector $n^\mu$, which represents the four-velocity of fiducial observers (FIDO). These provide the local frame with respect to which physical quantities are measured. The metric tensor and 3+1 temporal and spatial projection operators consist of, respectively:
\begin{equation}
    \label{eqSM:3+1proj}
    ds^2 = -\alpha^2 dt^2 + \gamma_{ij} (dx^i + \beta^i dt)(dx^j + \beta^j dt) \ ; \qquad T^\mu_{\ \nu} = - n^\mu n_\nu \ ; \qquad S^\mu_{\ \nu} \equiv \gamma^\mu_{\ \nu} = \delta^\mu_{\ \nu} + n^\mu n_\nu \ ,
\end{equation}
where $\gamma_{ij}$ is the spatial hypersurface metric.

The generalized parallel four-velocity is $\UE^\mu = {\Pi_\mathcal{E}}^\mu_{ \ \nu} U^\nu$, where the electromagnetic eigenspace projectors are defined as:
\begin{equation}
    \label{eqSM:projectors}
     \Pi_\mathcal{E}{}^\mu_{\ \nu} = \frac{F^\mu_{\ \alpha} F^\alpha_{\ \nu}+\mathcal{B}^2\delta^\mu_{\ \nu}}{\mathcal{E}^2+\mathcal{B}^2} \ ; \quad
     \Pi_\mathcal{B}{}^\mu_{\ \nu} = \frac{\mathcal{E}^2\delta^\mu_{\ \nu}-F^\mu_{\ \alpha} F^\alpha_{\ \nu}}{\mathcal{E}^2+\mathcal{B}^2} \ .
\end{equation} 
Applying the parallel projector definition to the guiding-center four-velocity, and employing the property $F^\mu_{\ \nu} U^\nu = (\bvec{U}\cdot\bvec{E})n^\mu + (\gamma_U\bvec{E}+\bvec{U}\times\bvec{B})^\mu$, for any vector $U^\mu = \gamma_U n^\mu + \bvec{U}^\mu$, results in:
\begin{align}
    \label{eqSM:projUE}
    \UE^\mu = \frac{(B^2+\mathcal{E}^2)\gamma_U - (\bvec{E}\times\bvec{B})\cdot\bvec{U}}{\mathcal{E}^2+\mathcal{B}^2} n^\mu + \frac{(\mathcal{B}^2-B^2)U^\mu + E^\mu(\bvec{E}\cdot\bvec{U}) + B^\mu(\bvec{B}\cdot\bvec{U}) + \gamma_U (\bvec{E}\times\bvec{B})^\mu}{\mathcal{E}^2+\mathcal{B}^2} \ .
\end{align} 
The temporal part of Eq. (\ref{eqSM:projUE}) is by definition the guiding-center Lorentz factor $\gamma = -n_\mu \UE^\mu$, and can be factored as $\gamma = \gamma_E^2 (\gamma_U - \bvec{v}_E\cdot\bvec{U})$, where $\gamma_E^2= 1/(1-v_E^2) = \frac{B^2+\mathcal{E}^2}{\mathcal{B}^2+\mathcal{E}^2}$ is the Lorentz factor of the purely spatial electric drift velocity $\bvec{v}_E$, which is defined as: 
\begin{equation}
     \label{eqSM:vE}
     \bvec{v}_E = \frac{\bvec{E}\times\bvec{B}}{B^2+\mathcal{E}^2} = \frac{\bvec{E}\times\bvec{B}}{\gamma_E^2 (\mathcal{B}^2+\mathcal{E}^2)} \ .
\end{equation}
This compact formula is equivalent to the expression found in the classic literature, $\bvec{v}_E = \frac{\bvec{w}_E}{2w_E^2} \left(1-\sqrt{1-4w_E^2}\right)$, with $\bvec{w}_E = \frac{\bvec{E}\times\bvec{B}}{E^2+B^2}$ \cite{beklemishev1999S,bacchini2020S}. Employing $\sqrt{1-4w_E^2} = \frac{\mathcal{E}^2+\mathcal{B}^2}{E^2+B^2}$ and substituting $w_E^2 = \frac{E^2B^2-\mathcal{E}^2\mathcal{B}^2}{(E^2+B^2)^2}$ yields Eq. (\ref{eqSM:vE}). 
The identities $E^2 - B^2 = \mathcal{E}^2 - \mathcal{B}^2$ and $\bvec{E}\cdot\bvec{B} = \mathcal{E}\mathcal{B}$ were employed.

\medskip
The spatial part of Eq. (\ref{eqSM:projUE}) is simplified by using the triple cross product identity $(\bvec{E}\times\bvec{B})(\bvec{E}\times\bvec{B}) = (\bvec{E}\times\bvec{B})^2 I - B^2\bvec{E}\bvec{E} - E^2\bvec{B}\bvec{B} + (\bvec{E}\cdot\bvec{B})(\bvec{E}\bvec{B}+\bvec{B}\bvec{E})$, where $I$ denotes the identity matrix, in its first term. This leads to:
\begin{align}
    \label{eqSM:projUE2}
     \frac{(\mathcal{B}^2-B^2)\bvec{U} + \bvec{E}(\bvec{E}\cdot\bvec{U}) + \bvec{B}(\bvec{B}\cdot\bvec{U})}{\mathcal{E}^2+\mathcal{B}^2} 
     = \frac{(\mathcal{B}\bvec{B}+\mathcal{E}\bvec{E})(\mathcal{B}\bvec{B}+\mathcal{E}\bvec{E}) - (\bvec{E}\times\bvec{B})(\bvec{E}\times\bvec{B})}{(B^2+\mathcal{E}^2)(\mathcal{B}^2+\mathcal{E}^2)}\cdot\bvec{U} \ ,
\end{align} 
with $(\bvec{E}\times\bvec{B})^2 = (B^2-\mathcal{B}^2)(B^2+\mathcal{E}^2)$. Since the remaining term of the spatial part of Eq. (\ref{eqSM:projUE}) is $\gamma_U\frac{\bvec{E}\times\bvec{B}}{\mathcal{E}^2+\mathcal{B}^2} = \gamma_E^2\gamma_U\bvec{v}_E$, we identify the generalized parallel magnetic field direction as:
\begin{align}
    \label{eqSM:bstar}
     \bvec{b}_* = \frac{\mathcal{B} \bvec{B} + \mathcal{E}\bvec{E}}{\gamma_E (\mathcal{B}^2+\mathcal{E}^2)} 
     \implies \bvec{U}_\mathcal{E} = \bvec{b}_*(\bvec{b}_*\cdot\bvec{U}) + \gamma\bvec{v}_E \ .
\end{align} 
This finally implies that the generalized parallel four-velocity $\UE^\mu$ is given by:
\begin{align}
    \label{eqSM:UEfinal}
    \UE^\mu = \gamma (n^\mu + v_E^\mu) + u_\parallel b_{*}^\mu \ ,
\end{align} 
where $u_\parallel = U^\nu b_{*\nu}$ is the parallel guiding-center velocity component, and the guiding-center Lorentz factor is defined at lowest order from the condition $U_\mathcal{E}^\mu {U_\mathcal{E}}_\mu = -1-2\mu\mathcal{B}/m$, such that $\gamma = \gamma_E\sqrt{1+u_\parallel^2+2\mu\mathcal{B}/m}$.
We have verified analytically that the formula for $\bvec{b}_*$ in Eq. (\ref{eqSM:bstar}) is equivalent to the expressions found in Refs. \cite{beklemishev1999S,bacchini2020S}.

\subsection*{3+1 decomposition of the perpendicular guiding-center four-velocity}

The identification of the canonical particle drifts requires the 3+1 spatial projection of the perpendicular guiding-center four-velocity (\ref{eq:UBcovariant}), which is compactly written as 
\begin{equation}
    \label{eqSM:UBcompact}
     \UB^\mu = -\frac{m}{q\mathcal{B}^2}{\Pi_\mathcal{B}}^\mu_{\ \alpha}F^\alpha_{\ \nu} \left( \UE^\lambda \nabla_\lambda \UE^\nu + \frac{\mu}{m}\nabla^\nu\mathcal{B} \right) \ .
\end{equation}
The decomposition (\ref{eqSM:bstar}) yields directly that the $\mathcal{E}$-eigenspace projector satisfies ${\Pi_\mathcal{E}}^\mu_{ \ \nu} U^\nu = -(U^\nu u_{E\nu}) u_E^\mu + (U^\nu b_{*\nu})b_{*}^\mu$, with $u_E^\mu = \gamma_E (n^\mu + v_E^\mu)$ the electric drift four-velocity. The electromagnetic projection operators (\ref{eqSM:projectors}) can therefore be cast in the equivalent forms:
\begin{equation}
    \label{eqSM:projectorsuE}
     \Pi_\mathcal{E}{}^\mu_{\ \nu} = -u_E^\mu u_{E\nu} + b_*^\mu b_{*\nu} \ ; \qquad
     \Pi_\mathcal{B}{}^\mu_{\ \nu} = \delta^\mu_{\ \nu} + u_E^\mu u_{E\nu} - b_*^\mu b_{*\nu} \ ,
\end{equation}
which greatly simplify the 3+1 decomposition of Eq. (\ref{eqSM:UBcompact}). 
We note that the projectors commute with the electromagnetic tensor, i.e., ${\Pi_\mathcal{B}}^\mu_{\ \nu} F^\nu_{\ \alpha} = F^\mu_{\ \nu} {\Pi_\mathcal{B}}^\nu_{\ \alpha}$.
Moreover, for simplicity of notation, we define $\mathcal{F}^\nu \equiv \UE^\lambda \nabla_\lambda \UE^\nu + \frac{\mu}{m}\nabla^\nu\mathcal{B}$, such that applying the perpendicular projector (\ref{eqSM:projectorsuE}) to Eq. (\ref{eqSM:UBcompact}) gives:
\begin{align}
    \label{eqSM:UBprojcovariant}
     {\Pi_\mathcal{B}}^\mu_{\ \alpha}F^\alpha_{\ \nu} \mathcal{F}^\nu &= 
     (\bvec{\mathcal{F}}\cdot\bvec{E})n^\mu + (\mathcal{F}^n\bvec{E} + \bvec{\mathcal{F}}\times\bvec{B})^\mu \\
     &+ u_E^\mu \left[ -\gamma_E(\bvec{\mathcal{F}}\cdot\bvec{E}) + \gamma_E (\mathcal{F}^n\bvec{E} + \bvec{\mathcal{F}}\times\bvec{B})\cdot\bvec{v}_E \right] 
     - b_*^\mu \left[ (\mathcal{F}^n\bvec{E} + \bvec{\mathcal{F}}\times\bvec{B})\cdot\bvec{b}_* \right] \nonumber \ ,
\end{align}
where $\mathcal{F}^n \equiv -n_\mu \mathcal{F}^\mu$ denotes the temporal projection of $\mathcal{F}^\mu$.

The temporal part of $U_\mathcal{B}^\mu$ satisfies $\gamma_\mathcal{B} \equiv -n_\mu \UB^\mu = \bvec{v}_E\cdot\bvec{U}_\mathcal{B}$, and therefore the spatial part $\bvec{U}_\mathcal{B}$ uniquely determines the structure of the entire four-vector.
As such, the spatial projection of Eq. (\ref{eqSM:UBprojcovariant}) is given by:
\begin{equation}
    \label{eqSM:UBspatial}
     \bvec{U}_\mathcal{B} = -\frac{m}{q\mathcal{B}^2} \Big[ \mathcal{F}^n\bvec{E} + \bvec{\mathcal{F}}\times\bvec{B} + \gamma_E^2 \bvec{v}_E \left[ (\bvec{\mathcal{F}}\times\bvec{B})\cdot\bvec{v}_E -\bvec{\mathcal{F}}\cdot\bvec{E} \right] 
     - \bvec{b}_* \left[ (\mathcal{F}^n\bvec{E} + \bvec{\mathcal{F}}\times\bvec{B})\cdot\bvec{b}_* \right] \Big] \ .
\end{equation}
This equation can be reduced to a simpler form by employing the parallel and perpendicular decomposition of the electric and magnetic fields in relation to $\bvec{b}_*$,
\begin{equation}
    \label{eqSM:EBperppar}
    \bvec{E} = (\bvec{E}\cdot\bvec{b}_*)\bvec{b}_* + (\bvec{B}\cdot\bvec{b}_*)\bvec{b}_*\times\bvec{v}_E \ ; \qquad
    \bvec{B} = (\bvec{B}\cdot\bvec{b}_*)\bvec{b}_* - (\bvec{E}\cdot\bvec{b}_*)\bvec{b}_*\times\bvec{v}_E \ ,
\end{equation}
which, inserted into Eq. (\ref{eqSM:UBspatial}), finally yields:
\begin{equation}
    \label{eqSM:UBspatialfinalF}
     \bvec{U}_\mathcal{B} = \frac{m}{q\mathcal{B}^2} (\bvec{B}\cdot\bvec{b}_*) \bvec{b}_*\times \left( \bvec{\mathcal{F}} - \mathcal{F}^n \bvec{v}_E \right) \ .
\end{equation}

The temporal and spatial projections of $\mathcal{F}^\nu$ are determined from the 3+1 decomposition of the geodesic equation. 
Splitting the guiding-center four-velocity as in Eq. (\ref{eqSM:UEfinal}), $\UE^\mu=\gamma n^\mu+\bvec{U}_\mathcal{E}^\mu$ with $\bvec{U}_\mathcal{E}=u_\parallel \bvec{b}_*+\gamma \bvec{v}_E$ and $n_\mu\bvec{U}_\mathcal{E}^\mu=0$, yields:
\begin{equation}
    \label{eqSM:DUEDt}
     \UE^\mu \nabla_\mu \UE^\nu=\left( \frac{d\gamma}{d\tau} - \gamma \bvec{g}\cdot\bvec{U}_\mathcal{E} - K\cdot\bvec{U}_\mathcal{E}\cdot\bvec{U}_\mathcal{E} \right)n^\nu
     +\left( \frac{DU_\mathcal{E}^i}{D\tau}\bvec{e}_i - \gamma^2\bvec{g} - \gamma H\cdot\bvec{U}_\mathcal{E} \right)^\nu \ ,
\end{equation}
where the spatial material covariant derivative is 3+1 decomposed as:
\begin{align}
    \label{eqSM:3+1DDt}
     \frac{D \UE^i}{D\tau} & \equiv \gamma^i_{\ \nu} \UE^\mu\nabla_\mu \UE^\nu + \gamma (H-K)^i_{\ j}\,\UE^j \nonumber \\ &= \frac{\gamma}{\alpha}\partial_t \UE^i + \left(\UE^j - \frac{\gamma}{\alpha}\beta^j \right) \partial_j \UE^i + \Gamma^i_{jk} \UE^j\left(\UE^k -\frac{\gamma}{\alpha}\beta^k \right) + \frac{\gamma}{\alpha}\gamma^{ij} (\partial_t \gamma_{jk}) \UE^k \ .
\end{align}

Finally, from Eq. (\ref{eqSM:DUEDt}), the temporal and spatial projections of $\mathcal{F}^\nu$ read:
\begin{align}
    \label{eqSM:gammaF}
     \mathcal{F}^n &= \frac{d\gamma}{d\tau} - \left( \frac{d\gamma}{d\tau} \right)_\text{GR} - \frac{\mu}{m}\,n^\mu\nabla_\mu\mathcal{B} \ , \\
    \label{eqSM:Fspatial}
     \bvec{\mathcal{F}} &= \frac{du_\parallel}{d\tau}\bvec{b}_* + u_\parallel\frac{Db_*^i}{D\tau}\bvec{e}_i + \frac{d\gamma}{d\tau}\bvec{v}_E + \gamma\frac{Dv_E^i}{D\tau}\bvec{e}_i - \gamma^2\bvec{g} - \gamma H\cdot\bvec{U}_\mathcal{E} + \frac{\mu}{m}\bvec{\nabla}\mathcal{B} \ ,
\end{align}
where $\left( \frac{d\gamma}{d\tau} \right)_\text{GR} \equiv \gamma\bvec{g}\cdot\bvec{U}_\mathcal{E} + K\cdot\bvec{U}_\mathcal{E}\cdot\bvec{U}_\mathcal{E}$.
Substituting the decompositions (\ref{eqSM:gammaF})-(\ref{eqSM:Fspatial}) into Eq. (\ref{eqSM:UBspatialfinalF}) and using the property $\bvec{B}\cdot\bvec{b}_* = \gamma_E \mathcal{B}$ leads to the final form of the generalized perpendicular four-velocity:
\begin{equation}
    \label{eqSM:UBspatialfinal} 
     \bvec{U}_\mathcal{B} = \frac{m\gamma_E}{q\mathcal{B}} \bvec{b}_*\times \Bigg[ u_\parallel \frac{D b_*^i}{D\tau}\bvec{e}_i + \gamma \frac{D v_E^i}{D\tau}\bvec{e}_i + \frac{\mu}{m} \bvec{\nabla}\mathcal{B} + \frac{\mu}{\alpha m} \left( \frac{\partial\mathcal{B}}{\partial t} - \bvec{\beta}\cdot\bvec{\nabla}\mathcal{B} \right)\bvec{v}_E -\gamma^2\bvec{g} - \gamma H\cdot\bvec{U}_\mathcal{E} + \left( \frac{d\gamma}{d\tau} \right)_\text{GR}\bvec{v}_E \Bigg] \ ,
\end{equation}
\begin{equation}
     \label{eqSM:UBfinal}
     \UB^\mu = (\bvec{v}_E\cdot\bvec{U}_\mathcal{B})n^\mu + \bvec{U}_\mathcal{B}^\mu \ .
\end{equation}
For a non-stationary metric, $\partial_t\gamma_{ij}\neq0$, the normalization $\left|\bvec{b}_* \right| = 1$ no longer implies $b_{*i}\frac{Db_*^i}{D\tau}=0$, and the second term in Eq. (\ref{eqSM:Fspatial}) enters the parallel force equation (\ref{eq:3+1du//dt}) as the additional term $-u_\parallel\frac{Db_*^i}{Dt}\bvec{e}_i\cdot\bvec{b}_*$.
The extrinsic curvature tensor likewise generalizes to $K_{ij}=H_{(ij)}-\frac{1}{2\alpha}\partial_t\gamma_{ij}$.

\medskip
The 3+1 decomposition of the parallel and perpendicular guiding-center velocities (\ref{eqSM:UEfinal}) and (\ref{eqSM:UBfinal}) can thus be employed to derive the 3+1 guiding-center equations (\ref{eq:3+1du//dt}), (\ref{eq:dXdtdrifts}) and (\ref{eq:3+1dgammadt}), as described in the main text, with the zeroth-order time dilation relation $\frac{d\tau}{dt}=\frac{\alpha}{\gamma}$. 
We note that retaining the temporal component of the perpendicular four-velocity implies that the guiding-center Lorentz factor measured by the fiducial observers becomes $\gamma+\gamma_\mathcal{B}$, such that, to first order, the time dilation relation becomes:
\begin{equation}
    \frac{d\tau}{dt} = \frac{\alpha}{\gamma+\gamma_\mathcal{B}} \simeq \frac{\alpha}{\gamma}\left(1-\bvec{v}_E\cdot\frac{\bvec{U}_\mathcal{B}}{\gamma}\right) \ .
\end{equation}
This correction expresses the relativistic velocity addition of the first-order drifts with the electric drift, and solely rescales the rate at which the guiding center evolves without introducing additional drifts. 
Its magnitude is $|\bvec{v}_E\cdot\bvec{U}_\mathcal{B}/\gamma| \sim \varepsilon \gamma_E v_E$, such that, for $\varepsilon \ll 1$, it is non-negligible only in regions where $\gamma_E \to \infty$, when the guiding-center expansion itself approaches its limit of validity. 
At first order, this correction only affects the zeroth-order terms of the guiding-center equations: namely the parallel and electric drift velocity terms in Eq. (\ref{eq:dXdtdrifts}) and the parallel electric field terms in Eqs. (\ref{eq:3+1du//dt}) and (\ref{eq:3+1dgammadt}). 
The gravitational polarization drift is therefore not affected by this correction, and does not contribute to it either, since it is perpendicular to $\bvec{v}_E$.

\section*{Numerical validation of the guiding-center equations of motion}
\label{sm:validation}

In order to validate the guiding-center formalism, we implemented the guiding-center equations of motion in the general-relativistic parallelized particle pusher with radiation reaction detailed in Ref. \cite{joaquim2026S}, and compared the trajectories obtained for single-particle motion by direct integration of Eqs. (\ref{eq:GCmu}), (\ref{eq:3+1du//dt}), and (\ref{eq:dXdtdrifts}), with the corresponding full-orbit evolution, detailed in the End Matter. 
The comparison was performed for several initial conditions in Schwarzschild, Kerr and slowly rotating Kerr spacetimes, for uniform and dipolar field geometries. 
The guiding-center trajectories and magnetic moment decay agree with the full-orbit simulations in all tested cases, provided that the GC approximation is valid. 

The scale separation between the particle gyromotion and the macroscopic gravitational dynamics is quantified by the ratio of characteristic frequencies:
\begin{equation}
    \label{eqSM:scaleseparation}
    \frac{|\omega_c|}{c/r_G} \approx 1.2\times10^{14}\left(\frac{\mathcal{B}}{10^{12}\,\mathrm{G}}\right)\left(\frac{M}{1.4M_\odot}\right) \ ,
\end{equation}
where the electron charge and mass were assumed and $r_G = GM/c^2$ is the gravitational radius of a body of mass $M$. 

For typical parameters of compact objects \cite{NScatalogS,BHcatalogS}, the scale separation (\ref{eqSM:scaleseparation}) ranges from $\sim 10^{10}-10^{11}$ for millisecond pulsars and stellar-mass black holes, to $\sim 10^{13}-10^{14}$ in pulsars and supermassive black holes, reaching $\sim 10^{17}$ for magnetars.
Since full-orbit integration must resolve the gyration period measured by fiducial observers, $T_L = 2\pi\gamma/|\omega_c|$, and would thus require at least $(10^{10}-10^{17}) \, \gamma^{-1}$ timesteps per light-crossing time $r_G/c$, it becomes computationally unfeasible at realistic astrophysical parameters.

\begin{figure}[H]
    \centering
    \includegraphics[width=\linewidth]{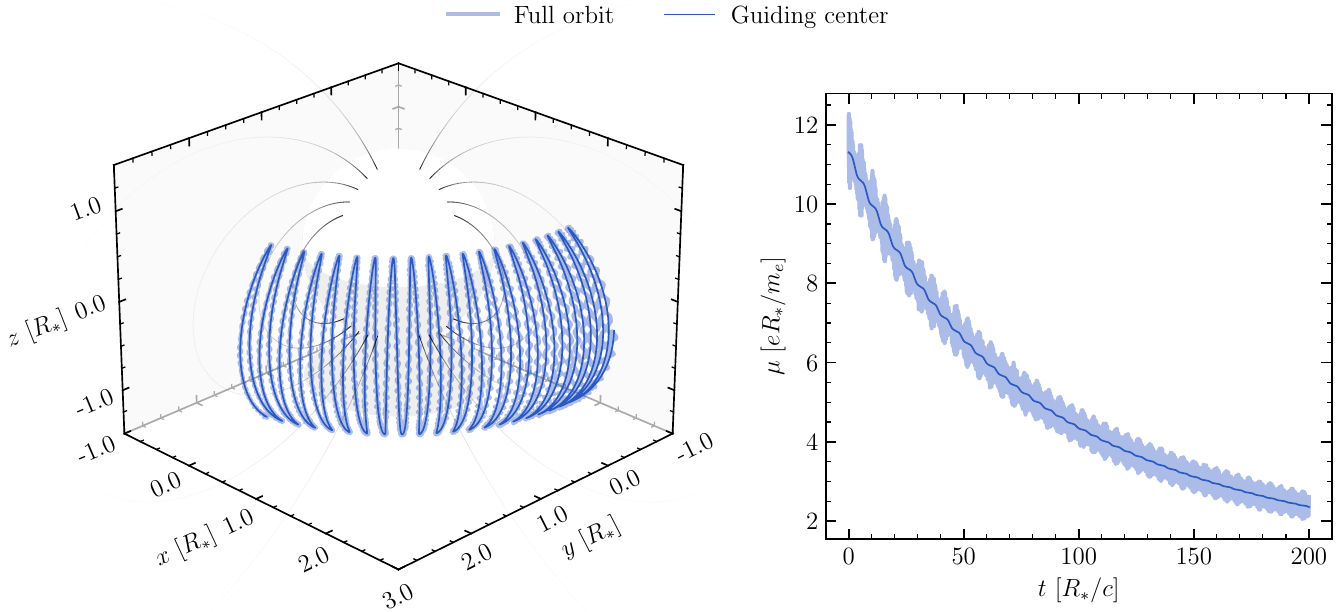}
    \caption{Comparison between the single-particle full-orbit and the guiding-center evolution for an electron initialized at $(x,y,z)=(3,0,0)\,R_*$ with $\gamma=10^3$ and a pitch angle $\theta_0 = -\pi/4$, subject to a magnetic dipole field in slowly rotating Kerr spacetime \cite{hartlethorne2S,rezzollafieldsS}, with stellar compactness $r_s/R_* = 0.5$ and stellar rotation $\Omega R_*/c = 0.1$, where $r_s$ and $R_*$ denote respectively the Schwarzschild and stellar radii. The simulation parameters are rescaled such that $|\omega_c|R_*/c = 10^{6}$ and $\omega_\text{RR}/|\omega_c| = 10^{-8}$ at the stellar pole. The agreement between the full-orbit and the guiding-center evolution validates the guiding-center system of equations.}
    \label{figSM:benchmark}
\end{figure}

In Fig. \ref{figSM:benchmark} we portray a representative case, with parameters chosen such that the drift velocities remain visible on the trajectory scale. We note that more intense fields reduce the expansion parameter $\varepsilon \propto \gamma/\mathcal{B}$ and therefore improve the agreement between the full-orbit and guiding-center trajectories, provided that the energy radiated per gyroperiod remains small.

\section*{Guiding-center particle trajectories in the M87${}^*$ jet}
\label{sm:M87jet}

Black-hole magnetospheres are a natural setting for the gravitational polarization drift. We apply the formalism of guiding-center theory with radiation reaction in curved spacetime to pair trajectories in the jet of the M87${}^*$ black hole, whose scale separation, $\left| \omega_c \right| r_G/c \sim 10^{13}$, from Eq. (\ref{eqSM:scaleseparation}), is beyond the reach of full-orbit simulations.

We model M87${}^*$ as a Kerr black hole in Boyer-Lindquist coordinates \cite{kerr1963S}, with mass $M = 6.5\times10^9 M_\odot$, gravitational radius $r_G = 9.6 \times 10^{14} \, \unid{cm}$, and spin $a=0.9$ \cite{eht2019S}, such that its event horizon lies at $r_H = r_G(1+\sqrt{1-a^2}) = 1.44 \, r_G$.
The black hole is threaded by the paraboloidal Blandford-Znajek (BZ) magnetospheric field with $B_* = 10\unid{G}$ at the horizon pole \cite{blandfordznajek1977S,eht2021S}, which is stationary, axisymmetric, and force-free.
The GC expansion parameter $\varepsilon \approx \gamma mc^2/(|q|\mathcal{B}r_G) \sim 10^{-13}\gamma$, taking the characteristic length scale of variation of the electromagnetic field as $L_B \approx r_G$, is the inverse of the scale separation.

\medskip
The electromagnetic field is determined from the magnetic flux function \cite{blandfordznajek1977S}:
\begin{equation}
    \label{eqSM:psiBZ}
    \Psi(r,\theta) = \Psi_0\left( r(1-\cos\theta) + 2r_G(1+\cos\theta)\left[1-\ln(1+\cos\theta)\right] \right) \ ,
\end{equation}
and the field lines that cross the ergosphere rotate rigidly with angular velocity $\Omega_F = \Omega_H/2$, where $\Omega_H = \frac{a}{2r_H}$ is the angular velocity of the horizon, while those that do not cross the ergosphere have $\Omega_F = 0$ and $B^{\hat\phi} = 0$ \cite{komissarov2004BHelectrodynamicsS,parfreyjetsS}.
In the fiducial orthonormal frame, the electric and magnetic fields are given by:
\begin{equation}
    \label{eqSM:BZparabFIDO}
    B^{\hat r} = \frac{1}{\sqrt{\mathcal{A}}\sin\theta}\frac{\partial\Psi}{\partial \theta} \ ; \qquad B^{\hat\theta} = -\frac{\sqrt{\Delta}}{\sqrt{\mathcal{A}}\sin\theta}\frac{\partial\Psi}{\partial r} \ ; \qquad B^{\hat\phi} = -v_F B_p \ ; \qquad \bvec{E} = -\bvec{v}_F\times\bvec{B} \ ,
\end{equation}
where $B_p = \sqrt{(B^{\hat r})^2 + (B^{\hat\theta})^2}$ is the poloidal magnetic field, and the Kerr metric functions are $\Delta = r^2 - 2r_Gr + a^2r_G^2$, $\Sigma = r^2 + a^2r_G^2\cos^2\theta$, and $\mathcal{A} = (r^2+a^2r_G^2)^2 - a^2r_G^2\Delta\sin^2\theta$ \cite{kerr1963S,bardeen1972S}. The velocity of the field lines is $\bvec{v}_F = \frac{\rho}{\alpha}(\Omega_F - \omega_\text{FD}) \bvec{e}_{\hat\phi}$, where $\rho = \sqrt{\mathcal{A}/\Sigma}\,\sin\theta$ is the cylindrical radius, $\alpha = \sqrt{\Sigma\Delta/\mathcal{A}}$ the lapse function, and $\omega_\text{FD} = 2ar_G^2r/\mathcal{A}$ is the frame-dragging angular velocity.
The electric drift therefore satisfies $v_F = \gamma_E v_E$, such that Eq. (\ref{eq:dgammadtGP}) reduces to $(d\gamma/dt)_\text{GP} = v_F^2\,(d\gamma/dt)_\text{GR}$: the gravitational polarization drift outpowers gravity outside the light cylinder, $v_F > 1$.

\medskip

\begin{figure}[H]
    \centering
    \includegraphics[width=0.99\linewidth]{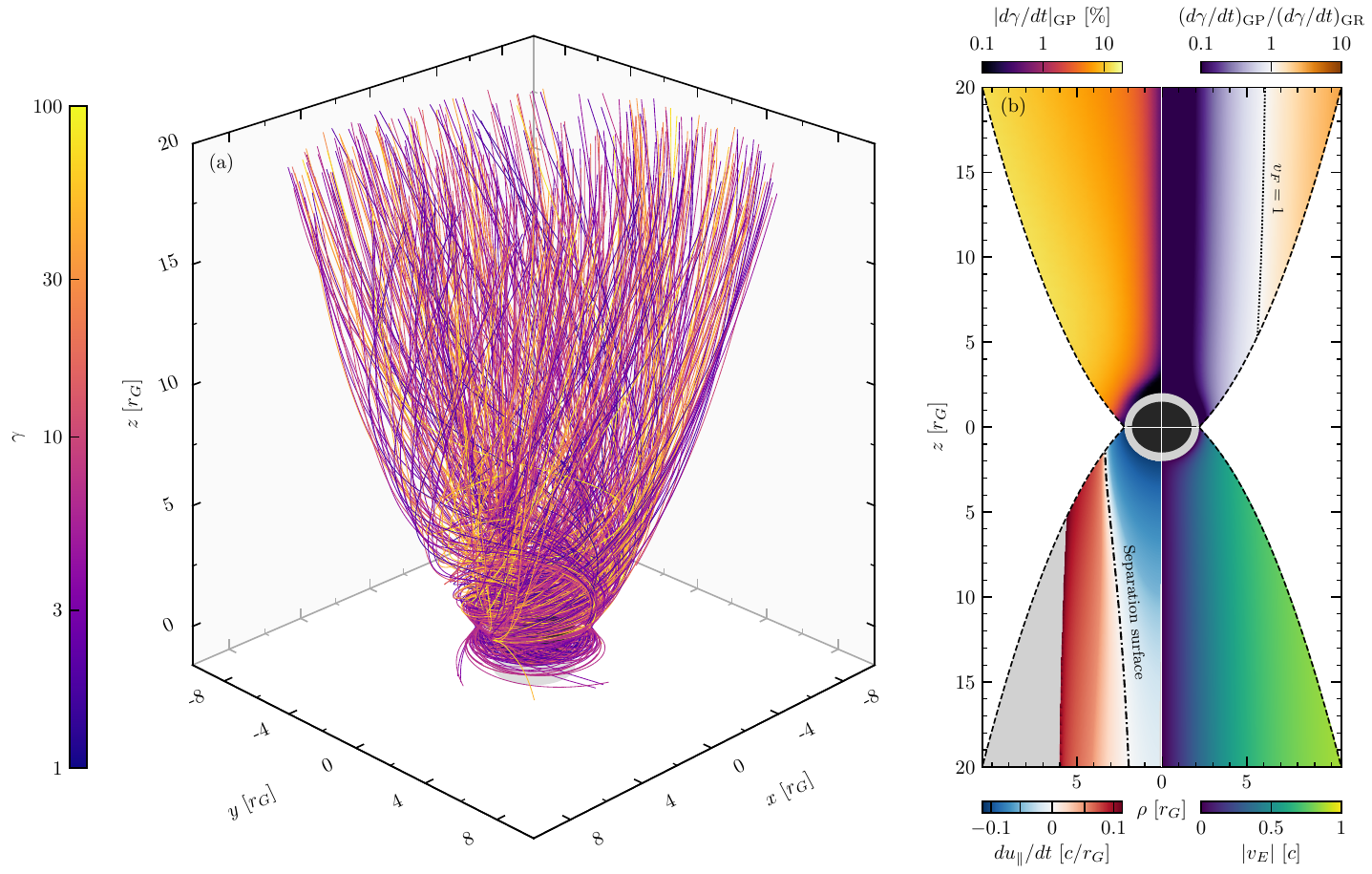}
    \caption{Electron-positron pair evolution in the jet of M87${}^*$, modeled as a Kerr black hole with spin $a=0.9$ threaded by a paraboloidal Blandford-Znajek magnetospheric field with $B_* = 10 \unid{G}$ at the horizon pole. The particle trajectories, colored by Lorentz factor, are depicted in the three-dimensional panel (a). The meridional plane (b) shows the fractional contribution of the gravitational polarization drift to the energy exchanged by the pairs (top left), and its comparison with the power of the gravitational force (top right); and the parallel acceleration of cold particles at rest on the rotating field lines, which determines the separation surface (dash-dotted line, bottom left), and the magnitude of the electric drift velocity (bottom right). The black dashed line is the last field line that crosses the ergosphere and the dotted line is the light cylinder $v_F = 1$. The regions where cold particles cannot corotate, and where particles are not injected, $r < 2\,r_G$, are represented in gray.}
    \label{figSM:M87}
\end{figure}

Following Refs. \cite{moscibrodzka2011S,wong2021S}, we inject particles on the rotating field lines, reproducing pair creation by photon-photon collisions of the MeV photons emitted by the accretion flow, at a rate per unit proper volume $\propto r^{-6}$ for $2 < r/r_G < 20$ independent of the polar angle. The injection is isotropic in the fiducial frame with an initial energy $\ln\gamma_0$ uniformly distributed in $[0,\ln 100]$ \cite{moscibrodzka2011S}.
Inside the null surface $\omega_\text{FD} = \Omega_F$, at $r \lesssim 2\,r_G$, $v_F$ and $B^{\hat\phi}$ change sign, and we therefore restrict the injection to $r > 2\,r_G$.
Since electrons and positrons follow the same trajectories up to kinematically negligible first-order drifts, and have the same energization rates, in Fig. \ref{figSM:M87} we evolve $N=10^6$ electrons for $200\,r_G/c$, and show the resulting trajectories and energization rate due to the gravitational polarization drift.
%
%
%
Of the injected pairs, $35\%$ escape along the jet, while the captured pairs fall into the black hole within $\sim5\,r_G/c$ (median): $95\%$ of infalling pairs start at $r_0 < 3\,r_G$. 
A small fraction of the pairs initialized near the equatorial plane at $r_0 \approx 2$-$3\,r_G$ on the outermost rotating field lines, $\approx 2\%$, remain trapped there, corotating with the field lines, as seen at the base of panel (a).
The escaping pairs, $86\%$ of which start with $u_\parallel > 0$, are first dragged azimuthally by the rotating field lines near the base of the jet, and then stream along it.
For cold particles initially at rest on the rotating field lines, the guiding-center equations yield a nearly cylindrical separation surface at $\rho \approx 2.5-3.5\,r_G$.
These values are comparable with the shape of the jet stagnation surface and with the range $r \approx 5-10 r_G$ found in GRMHD simulations \cite{nakamura2018separationsurfaceS}.

The power delivered by the gravitational polarization drift exceeds that of the gravitational force outside the light cylinder $v_F = 1$. Its share of the energy exchanged by the pairs grows across the jet, reaching $\sim 13\%$ at the jet edge at $z = 20\,r_G$.
When the gravitational polarization drift is accounted for, Eq. (\ref{eq:3+1dgammadt}), integrated along the trajectories of the outflowing pairs, reproduces the simulated change in energy to within $0.04\%$ (median), and differs from it by $44\%$ (median) when the drift is not included.
The gravitational polarization drift is therefore a significant energization channel in black-hole jets.

\makeatletter\global\let\@FMN@list\@empty\makeatother

\end{document}